\documentclass[prd,twocolumn,amsmath,amssymb,superscriptaddress,floatfix,nofootinbib]{revtex4}
\usepackage{mathrsfs,bm}
\usepackage{longtable,lscape}
\usepackage{txfonts}
\usepackage{amssymb}
\usepackage{indentfirst}
\usepackage{graphicx,,booktabs}
\usepackage{multirow, overpic}
\usepackage{color}
\usepackage{slashed}
\usepackage{amssymb}

\usepackage{color}
\usepackage[bookmarks=true,bookmarksopen=false,plainpages=false,breaklinks=true,
   bookmarksnumbered=true,hypertexnames=false,
   filecolor=blue,urlcolor=three,menucolor=three,
   linkcolor=three,citecolor=blueone, colorlinks,
   anchorcolor=blue,runcolor=pink,frenchlinks=red
   pdfstartview=FitH,pdftitle=title,%
   pdfauthor=author]{hyperref}

\allowdisplaybreaks[4]

\definecolor{cover}{rgb}{0.77,0.87,0.88}
\definecolor{blueone}{rgb}{0.1,0.1,.7}
\definecolor{citec}{rgb}{0.14,0.47,0.09}
\definecolor{two}{rgb}{0.0,0.5,0.}
\definecolor{three}{rgb}{.5,.1,0.15}

\begin{document}

\title{New reaction approach to reflect exotic structure of hadronic molecular
state}

\author{Zuo-Ming Ding} \affiliation{School of Physics and Technology, Nanjing Normal University, Nanjing 210097, China}

\author{Jun He} \email{junhe@njnu.edu.cn (Corresponding author)}
\affiliation{School of Physics and Technology, Nanjing Normal University, Nanjing 210097, China} \affiliation{Lanzhou Center for Theoretical Physics, Lanzhou University, Lanzhou 730000, China}

\author{Xiang Liu} \email{xiangliu@lzu.edu.cn (Corresponding author)}
\affiliation{Lanzhou Center for Theoretical Physics, Lanzhou University, Lanzhou 730000, China} \affiliation{School of Physical Science and Technology, Lanzhou University, Lanzhou 730000, China} \affiliation{Key Laboratory of Theoretical Physics of Gansu Province, and Frontiers Science Center for Rare Isotopes, Lanzhou University, Lanzhou 730000, China} \affiliation{Research Center for Hadron and CSR Physics, Lanzhou University and Institute of Modern Physics of CAS, Lanzhou 730000, China}

\date{\today}
\begin{abstract}

With the accumulation of  experimental data, more and more exotic
hadrons are observed. Among the interpretations of  these exotic hadrons,
molecular state and compact multiquark are two of the most popular pictures.
However, it is still difficult to determine the structure of an exotic hadron.  In
this work, we propose a possible way to detect the internal structure of an
exotic state.  When a molecular state composed of two consistent hadrons is
attacked by another particle, one of the constituents should be kicked out while
another quasifree constituent stays almost unaffected. It is different from a
compact multiquark which has no obvious subcluster. In this work, taking the $X
(3872)$ as an example, we perform a Dalitz plot analysis of such reaction to find
the effect of the different internal structures. Under the assumption of the $X
(3872)$ as a molecular state or a compact tetraquark state, with the help of the
effective Lagrangians, the Dalitz plot and the invariant mass spectrum are
estimated with different total invariant mass of three final particles,  and the
effect of different binding energies is also discussed. Obvious event
concentration can be observed as strips in the Dalitz plot and sharp peaks in the
invariant mass spectrum for the $X(3872)$ with a small binding energy under the
molecular state picture, while such concentration cannot be observed under
the compact tetraquark  picture. Such phenomenon can be applied to identify the
internal structure of a new hadron state.

\end{abstract}

\maketitle

\section{Introduction}

The study of exotic hadrons is one of the most important topics in hadron
physics.  Theoretically, the basic theory of the strong interaction, quantum
chromodynamics (QCD), allows the existence of exotic hadrons beyond the
conventional picture where the hadrons are composed of  three quarks or a
quark-antiquark pair.  Experimentally, with the development of the experimental
techniques and accumulation of data, more and more exotic particles are observed
but cannot be put into the frames of the conventional quark
model~\cite{ParticleDataGroup:2018ovx,Godfrey:1985xj,Capstick:1986ter}.  If we
deem these new particles as a genuine state composed of quarks, there exist two
main interpretations: compact multiquark and hadronic molecular state. For most
of the exotic states, both interpretations exist simultaneously in the
literature~\cite{Chen:2016qju}.  It is an interesting and difficult problem to
determine the real internal structure of an exotic hadron.

The molecular state is a loosely bound state of two hadrons~\cite{Guo:2017jvc}.
Such idea is considerably easy understand if we take the deuteron as a
molecular state composed of two hadrons, that is, a nucleon.  It is also natural
to expect the existence of bound states from other hadrons. The experimental
observation seems to support such assumptions also. Considerable $XYZ$ particles
are close  to the thresholds of two hadrons~\cite{ParticleDataGroup:2018ovx}.
The most popular interpretation about such phenomenon is that these particles
are composed of the corresponding hadrons with a small binding energy as
deuteron~\cite{Chen:2016qju,Guo:2017jvc}.  The hadronic molecular state is, in
fact, a picture in the hadron level. Different from the molecular state, the compact
multiquark is a real bound state of quarks~\cite{Liu:2019zoy}. In a compact
multiquark, no obvious subcluster can be found, and  its radius is usually
assumed to be much smaller than a molecular state. Theoretically, the mass of
a compact multiquark is irrelevant to the thresholds of hadrons.  It seems to be
used to judge whether an exotic hadron is a molecular state. However,
practically, due to the uncertainty of both theory and experiment, many exotic
states near thresholds can be also explained as a compact multiquark. Moreover,
the multiquark is still an important picture to explain the states which are far
from any threshold.   

In the literature, the masses and the decay patterns  are the most important
ways to detect the internal structure of an exotic state. However, as said
above, the accordance of the theoretical mass with the experimental mass is not
enough to determine the internal structure of an exotic state because it often
can be explained in both pictures. The decay pattern is most promising to
reflect the quark distributions in the exotic hadrons. However, the
uncertainties from both experiment and theory make it difficult to reach a
determinative conclusion.  Hence, it is helpful to find more ways to detect the
internal structure of the exotic states.

The main difference between a molecular state and a  compact multiquark is the
spatial distribution of the quarks. In a molecular state, the quarks are grouped
into two hadrons that have a distance of  more than 10 fm.
In addition to effects on the decay pattern, such structure should be reflected when
being attacked by a particle. For a molecular state, the incoming particle,
which is usually about 1~fm can be easily attacked into the molecular state and even
pass through it.  Since the distance of two constituent hadrons is separated by
long distances, the collision happens only on one constituent. When a
constituent hadron is attacked, another one should be little affected.
However, for a compact multiquark, which is usually about 1~fm, the results of
the collision of an incoming particle is to excite the multiquark and induce its
decay. Because of  the compactness, the momenta of the incoming particle should be
transferred to all quarks and then all final particles. Hence, the molecular
state should have quite different behavior after collision.

Such difference of the behavior of collision should be a promising way to detect
the internal structure of an exotic state. However, there is an obvious difficulty.
We do not have enough stable exotic states to make a target or a beam to perform
a collision with another particle.  The direct measurement of such collision is
impossible with the current or near future experimental technology.  However,
such collisions can happen in the production of the exotic hadrons in a
nucleon-rich environment, which is the realistic scene at facilities, such as
LHCb and PANDA. If we can  extract the information of  collision of the nucleon
with produced exotic states, it is still promising to obtain enough events
to study such  different behaviors of the molecular state and compact multiquark. 

In the current work, we try to propose a scheme to realize such idea with the
well-known exotic particle $X(3872)$ as an example.  Despite being studied by hundreds of
experimental physicists and theorists, the structure of $X (3872)$ is not
yet fully understood. The most preferred interpretation of the structure of $X
(3872)$ nowadays is  $c\bar{c}-D{\bar{D}^*}$ mixing
state \cite{Badalian:2012jz,Wang:2010ej}. There also exist other interpretations, such
as pure molecular state~\cite{Belle:2003nnu,Tornqvist:2004qy,Swanson:2003tb},
compact tetraquark structure ($c\bar{c}q\bar{q}$) of this exotic
state \cite{LHCb:2020sey,Esposito:2021vhu}, radial excitation of the $P$-wave
charmonium \cite{Barnes:2005pb}, and a vector glueball mixed with neighboring
vector states of charmonium \cite{Seth:2004zb}.  Among these interpretations,
except for the nongenuine particle explanations such as triangular singularities,
the $X(3872)$ is a compact quark pair or tetraquark, or a molecular
state, or their mixing. In this work, we will study the behavior of the
$X(3872)$ attacked by a nucleon in two pictures, the compact quark state and
molecular state, to study collision of the $X(3872)$ with a proton as $p+X(3872)\to
p+\bar{D}^0+D^0$. 

This article is organized as follows. We present the
theoretical formalism to study the reaction of the $p+X(3872)\to
p+\bar{D}^0+D^0$ in two pictures in Section~\ref{THEORETICAL FRAME}.  The
numerical results  will be given in Section~\ref{NUMERICAL RESULTS}. Finally,
the article ends with a summary in section~\ref{Summary}.

\section{Collision of the $X(3872)$ with a proton} \label{THEORETICAL FRAME} 

\begin{figure}[h!]
	\scalebox{0.7}{\includegraphics[bb=60 0 900 550, clip, scale=0.42]{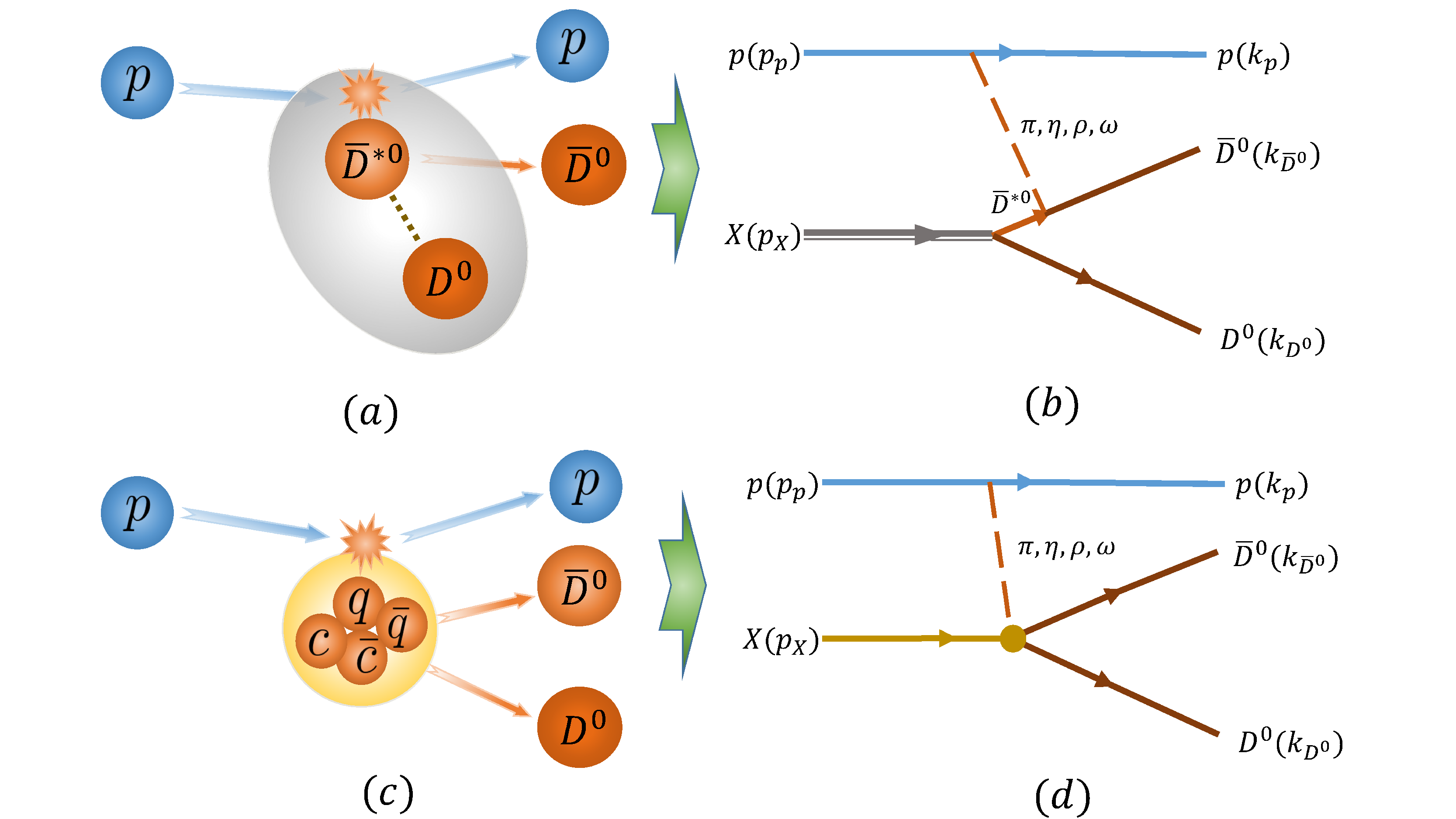}}
  \caption{The sketch map (a), (c) and Feynman diagram (b), (d)  of reaction $p+X(3872)\to p+\bar{D}^0+D^0$ with assumption  of  the $X (3872)$ as a loosely bound molecular state (a), (b)  or a compact multiquark state (c), (d).  The denotations of the momenta of particles are also given.}
  \label{diagram}
\end{figure}

In the current work, we will consider the process  $p+X(3872)\to
p+\bar{D}^0+D^0$ to detect the internal structure of the $X(3872)$ with a nucleon.
The LHCb experiment definitively established that the $X(3872)$ has $J^{P C} =
1^{++}$ \cite{LHCb:2013kgk}, which means that, even if the $X(3872)$ is the
$D^{0}{\bar{D}^{*0}}$ molecular state, it cannot decay into a $D$ and a $\bar{D}$
meson due to the conservation of spin parity.  But collision by a proton may make
this process possible. In Ref.~\cite{He:2022rta}, we studied the nucleon-induced
fissionlike process of the $T_{cc}^+$. Under an assignment of the $T_{cc}^+$ as
a molecular state, when induced by a proton, the $T_{cc}^+$ could decay into a $D$
and $D$ pair. Since the $T_{cc}^+$ and $X(3872)$ have some similar features such as
the small binding energy and narrow width, one can legitimately predict that the
$X(3872)$ is possible to decay into a $D$ and $\bar{D}$ pair~\cite{He:2021smz}.
Furthermore, this reaction can be used to reveal underlying structures of the
$X(3872)$.  Here, we consider both pictures of the molecular state and compact quark
state as shown in Fig.~\ref{diagram}.  

In the hypothesis of the $X(3872)$ as a loosely molecular state with wave
function $\left( {\bar{D}^{*0} }{D^0}-D^{*0} \bar{D}^{0}\right) / \sqrt{2}$, its
radius is supposed to be about 10 fm~\cite{Liu:2008fh} (hereafter, we use the first
part of the wave function for explanations; the results for the second part can
be obtained analogously). The nucleon and the constituent $\bar{D}^{*0}$ and
$D^0$ mesons have a radius smaller than 1~fm.  Hence, the proton should attack on
one of the constituent mesons of the $X(3872)$. If we only consider the process with
three final particles, proton, $D$, and $\bar{D}$ meson, the proton should
attack on the $\bar{D}^{*0}$ meson, as shown in Fig.~\ref{diagram} (a). After
the $\bar{D}^{*0}$ meson is attacked, it is transformed to a $\bar{D}^{0}$ meson.
Since the $X(3872)$ is a loosely bound state, it forms a quasi-two-body
scattering, $p\bar{D}^{*0}\to p\bar{D}^0$. The final $\bar{D}^0$ meson should be
obviously affected by the energy transferred from the incoming proton and that
released from the transition of the $\bar{D}^{*0}$ to $\bar{D}^{0}$ meson. However,
another constituent of the $X(3872)$, $D^0$ meson, should be little affected
due to the weak binding. Such process can be rewritten as the Feynman diagram in
Fig.~\ref{diagram} (b).

If the $X(3872)$ is a compact quark state, the collision behavior is quite
different, as shown in Fig.~\ref{diagram} (c). The radius of the $X(3872)$ should be
smaller than 1~fm for both quark-antiquark pair $[c\bar{c}]$ and tetraquark
$[c\bar{c}q\bar{q}]$ explanations of the $X(3872)$ (in Fig.~\ref{diagram} (c) and
hereafter, we mainly adopt the tetraquark picture). Additionally, such state is
composed of quarks binding tightly.  In such picture, the $X(3872)$ is excited
by the collision of the proton. The excited $X(3872)$ decays to a $\bar{D}^0$
meson and a $D^0$ meson. The energy transferred from the incoming proton will be
acquired by both final mesons, which is different from the molecular state
picture. The Feynman diagram can be written as Fig.~\ref{diagram} (d).

The collision above is difficult to be performed due to lack of the $X(3872)$ target
or beam. Here, we propose to consider the produced $X(3872)$ in a nucleon-rich
environment. Instead of considering the initial proton and $X(3872)$, three
final particles, the proton, and $\bar{D}^0$  and  $D^0$ mesons can be collected with
certain total invariant mass obtained as
$W=\sqrt{s}=\sqrt{P^2}=\sqrt{(k_p+k_{\bar{D}^0}+k_{k^0})^2}$  with
$k_{p,\bar{D}^0, D^0}$ being the momenta of final particles, which is
independent of the coordinate frames.  Among these events, the Dalitz plot against
invariant masses $m_{pD^0}=\sqrt{(p_p+p_{D^0})^2}$ and
$m_{p\bar{D}^0}=\sqrt{(p_p+p_{\bar{D}^0})^2}$ can be obtained by selecting the
corresponding event. In such treatment, all observations are invariant. 

The laboratory frame with the static $X(3872)$ will be adopted to perform explicit
deduction and numerical calculation.  In this reference frame, the cross section
for the reaction $p+X(3872)\to p+\bar{D}^0+D^0$ reads as
\begin{eqnarray}
	d\sigma=\frac{1}{4[(p_p\cdot p_X)^2-m_p^2m_X^2]^{1/2}}\frac{1}{6}\sum_{\lambda_p\lambda_{X}\lambda'_p}|{\cal M}_{\lambda_p\lambda_{X},\lambda'_p}|^2d\Phi_3,\label{cross}
\end{eqnarray}
where  $p_{p,X}$ and $m_{p,X}$ are the momentum and mass of the incoming proton
or the $X (3872)$. Practically, the GENEV code in FAWL is adopted to generate the event of three-body final state by the Monte Carlo method, that is, the phase space 
$$R_3=(2\pi)^{5}d\Phi_3=\prod^3_i\frac{d^3k_i}{2E_i}\delta^4\left(\sum^n_ik_i-P\right),$$
where  $k_i$ and $E_i$ are the momentum and energy of final particle $i$. 
The mechanism  can be described by an amplitude ${\cal
M}_{\lambda_p\lambda_{X},\lambda'_p}$ with $\lambda$ being the helicity of the incoming proton, $X(3872)$, or final proton. It will be derived with the Feynman diagrams in Fig.~\ref{diagram}. Here the interaction between the proton and the $X(3872)$ is described by light meson exchange. It has the same form in two pictures and can be obtained with the help of effective Lagrangians, 
\begin{eqnarray}
 \mathcal{L}_{{\mathbb P}NN}&=&-\frac{g_{ {\mathbb
 P}NN}}{\sqrt{2}m_N} \bar{N}_b\gamma_5\gamma_\mu
 \partial_\mu{\mathbb P}_{ba}  N_a,\\
 \mathcal{L}_{\mathbb{V}NN}&=&-\sqrt{2}g_{\mathbb{V} NN}
 \bar{N}_b\bigg(\gamma_\mu+\frac{\kappa}{2m_N}\sigma_{\mu\nu}\partial^\nu\bigg){\mathbb{V}}
 _{ba}^\mu N_a,\label{vv}
\end{eqnarray}
where $\mathbb P$ and $\mathbb V$ are two-by-two pseudoscalar and vector matrices.
$N^T=(p,n)$  is a field for the nucleon. 
The coupling constants $g^2_{\pi NN}/(4\pi)=13.6$, $g^2_{\rho NN}/(4\pi)=0.84$,
$g^2_{\omega NN}/(4\pi)=20$ with $\kappa=6.1~(0)$ for the $\rho~(\omega)$ meson,
which are used in the Bonn nucleon-nucleon potential ~\cite{Machleidt:2000ge}
and meson productions in nucleon-nucleon collision~\cite{Cao:2010km,
Tsushima:1998jz,Engel:1996ic}. The $\eta$ exchange is neglected in the current
work due to the weak coupling of $\eta$ or $\phi$ mesons to nucleons as indicated in
many previous works \cite{Machleidt:2000ge,Cao:2010km}. Here, a factor
$f_i(q^2)=(m_i^2-\Lambda^2)/(q^2-\Lambda^2)$ is also introduced to the propagator of
each exchanged meson with cutoff $\Lambda=1$~GeV.  The left part of the
amplitudes in two pictures  is different, as given below.

In the molecular state picture as shown in Fig.~\ref{diagram} (a), the exchanged
light meson interacts with the $\bar{D}^{*0}$ meson in the $X(3872)$. In terms of
heavy quark limit and chiral symmetry, the corresponding Lagrangians have been
constructed in the literature as~\cite{Casalbuoni:1996pg}, \begin{eqnarray}
\mathcal{L}_{\mathcal{P}^*\mathcal{P}\mathbb{P}} &=&-
\frac{2g}{f_\pi}(\mathcal{P}^{}_b\mathcal{P}^{*\dag}_{a\lambda}+\mathcal{P}^{*}_{b\lambda}\mathcal{P}^{\dag}_{a})\partial^\lambda{}\mathbb{P}_{ba},\\
\mathcal{L}_{\mathcal{P}^*\mathcal{P}\mathbb{V}} &=&- 2\sqrt{2}\lambda{}g_V
v^\lambda\varepsilon_{\lambda\alpha\beta\mu}
(\mathcal{P}^{}_b\mathcal{P}^{*\mu\dag}_a +
\mathcal{P}_b^{*\mu}\mathcal{P}^{\dag}_a)%\nonumber\\&&\times
\partial^\alpha{}\mathbb{V}^\beta_{ba}, \end{eqnarray} where
${\mathcal{P}}^{(*)T} =(D^{(*)0},D^{(*)+})$ is the field for the $D^{(*)}$ meson. The parameters involved here were determined
in the literature as $g=0.59$, $\beta=0.9$, $\lambda=0.56$ GeV$^{-1}$,
$g_V=5.9$, and $f_\pi=132$~MeV~\cite{Casalbuoni:1996pg,Chen:2019asm}.

In the molecular state picture, the amplitude  ${\cal A}_{\lambda_{X},\lambda_{\bar{D}^{*0}}}$ for the split
of the $X (3872)\to \bar{D}^{*0}D^{0}$ for the first term of the wave function is ~\cite{He:2022rta}
\begin{align}
  \frac{{\cal A}_{\lambda_{X},\lambda_{\bar{D}^{*0}}}}{p^2-m^2_{\bar{D}^{*0}}}&\simeq-\frac{\sqrt{8m_{X}m_{\bar{D}^{*0}}m_{D^{0}}}}{m_{X}-m_{D^{0}}+m_{\bar{D}^{*0}}}\psi({\bm k}_3)\epsilon_{\lambda_{X}}\cdot\epsilon^*_{\lambda_{\bar{D}^{*0}}},
  \end{align}
where  $\lambda_{X}$ and $\lambda_{\bar{D}^{*0}}$ are helicities for the initial $X (3872)$ state and intermediate $\bar{D}^{*0}$ meson. The $p$ and $m_{\bar{D}^{*0}}$ are the momentum and mass of the intermediate $\bar{D}^{*0}$ meson.
Here $m_{X,\bar{D}^{*0},D^{0}}$ is the mass of the $X (3872)$,  $\bar{D}^{*0}$ and $D^{0}$. The $\epsilon_{\lambda_{X}}$ and $\epsilon_{\lambda_{\bar{D}^{*0}}}$ are the polarized vectors of the $X (3872)$ and $\bar{D}^{*0}$ meson, respectively. The wave function $\psi({\bm k})=\sqrt{8\pi/a}/({\bm k}^2+1/a^2)$ with normalization $\int d^3k/(2\pi)^3|\psi(k)|^2=1$~\cite{Voloshin:2003nt}. Scattering length $a=1/\sqrt{2\mu E_B}$ with the reduced mass $\mu=m_{\bar{D}^{*0}}m_{D^0}/(m_{\bar{D}^{*0}}+m_{D^0})$ with $E_B$ being the binding energy.   

Different from the molecular state, the $X(3872)$ has no obvious subcluster in
the compact quark  state picture. In the molecular state picture, the large
radius and distance between two constituents make the collision happen on one
of the constituents. If the $X(3872)$ is a compact binding state of quarks with
small radius, the proton should attack the $X(3872)$ in the whole, which will
be excited by the light meson emitted by the proton and decays into two $D$
mesons.  In the current work, we do not consider its explicit mechanism. However, such
interaction should happen in a small space and short time, which can be taken as
a four particle vertex as shown in Fig.~\ref{diagram} (d). Such vertex can
be written as effective Lagrangians, \begin{eqnarray}
\mathcal{L}_{X\mathcal{P}\mathcal{P}\mathbb{P}} &=&g_{X} X^{\mu} \partial_{\mu}
\mathbb P \mathcal{P} \mathcal{P}, \\
\mathcal{L}_{X\mathcal{P}\mathcal{P}\mathbb{V}} &=&g_{X}
\varepsilon_{\alpha\beta\zeta\eta} \partial^\alpha \mathbb{V}^\beta
\partial^\zeta X^{\eta} \mathcal{P} \mathcal{P}. \end{eqnarray} Since no
explicit mechanism is introduced, the coupling constant  $g_X$ for each exchange
cannot be determined, which will be discussed later.

\section{Numerical results}\label{NUMERICAL RESULTS}

\begin{figure}[htpb] \scalebox{1}{\includegraphics[bb=80 70 500 300,
clip,scale=1.5]{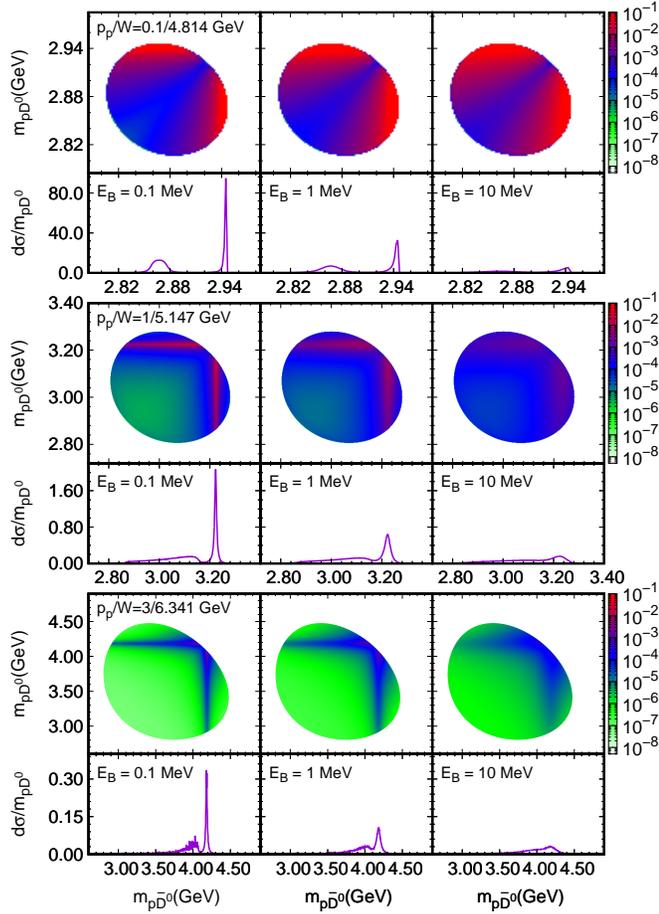}} \caption{Event distribution  for the $p+X(3872)\to
p+\bar{D^0}+D^0$ reaction assuming the $X (3872)$ as a loosely molecular state. The
momentum of the incoming proton or total invariant mass ${\rm p}_p/W$ = 0.1/4.814,
1/5.147 and 3/6.341~GeV, and the binding energy $E_B$ = 0.1, 1 and 10 MeV,
respectively. For each example choice of ${\rm p}_p/W$, the figures represent
the Dalitz plot $d\sigma/dm_{p\bar{D}^0}dm_{pD^0}$ in the
$m_{p\bar{D}^0}-m_{pD^0}$ plane in a bin of $\mu$b/0.002$\times$0.002~GeV (upper) and invariant mass spectrum $d\sigma/dm_{p\bar{D}^0}$ against $m_{pD^0}$
in a bin of $\mu$b/0.002 GeV (lower). The results are obtained with
$10^{11}$ simulations.}\label{Dalitz1} \end{figure}

Since the incoming momentum cannot be measured in the scene of the nucleon-rich
environment, we consider the process $p+X(3872)\to p+\bar{D}^0+D^0$ with a total
invariant mass $W=\sqrt{s}$. With certain $W$, the event distribution can be
obtained as the Dalitz plot and invariant mass spectrum against $m_{pD^0}$ and
$m_{p\bar{D}^0}$.  The different internal structures of the $X(3872)$ will
be exhibited in the Dalitz plot and invariant mass spectrum.  The experimental
binding energy of the $X(3872)$ is every small and even above the thresholds. In
the current work, we would like to take it as an example to explain how to
detect the internal structure of an exotic hadron. Hence, different values of
binding energy will be adopted to show the variation of the Dalitz plot and
spectrum with the binding energy. The results in the molecular state picture are
shown in Fig.~\ref{Dalitz1}, where the Dalitz plot in the 
$m_{p{D}^0}$-$m_{p\bar{D}^0}$ plane for the reaction $p+X (3872) \to
p+\bar{D}^{0}+D^0$ is calculated with different binding energies of the $X (3872)$
as $E_B$=0.1, 1, and 10 MeV.  Correspondingly, the invariant mass  spectrum
against $m_{p\bar{D}^0}$ is also presented under each Dalitz plot. The momentum
of the incoming proton ${\rm p}_p=|{\bm p}_p|$ will affect the event distributions and is
chosen as 0.1, 1 and 3 GeV, which corresponds to total invariant mass $W=$ 4.814, 5.147, and
6.341~GeV, respectively.

Because of the symmetry in the wave function, analogical distributions for
$m_{p\bar{D}^{0}}$ and $m_{pD^0}$ can be observed.  For example, the Dalitz plot
in the first column of the third row in Fig.~\ref{Dalitz1}, with an incoming
momentum or total invariant mass ${\rm p}_p/W$ = 3/6.341 GeV, obvious and
similar strips can be found at both $m_{p\bar{D}^0}$ and $m_{p{D}^0}$ of about
4.19 GeV in the Dalitz plot.  With the elevation of the incoming proton ${\rm p}_p/W$, the
phase space of the final states becomes larger. The area of Dalitz plots in the 
$m_{p{D}^0}$-$m_{p\bar{D}^0}$ plane for ${\rm p}_p/W$=0.1/4.814 GeV  is much
smaller than that for 3/6.341 GeV. Hence, the results are rescaled; that is,
different ranges are adopted for three momenta.

\begin{figure}[h!] \includegraphics[bb=72 70 220 300,
clip,scale=1.6]{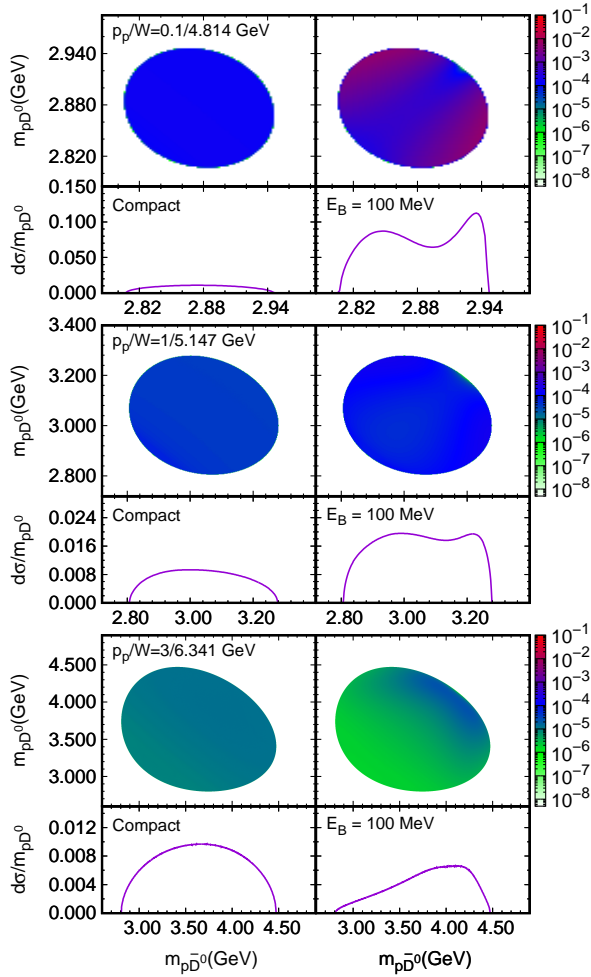} \caption{ Event distribution for the $p+X(3872)\to
p+\bar{D}^0+D^0$ reaction under the assumption of the $X (3872)$ as a compact quark
state in the first column  and under the assumption of the $X (3872)$ as a molecular
state with binding energy $E_B$ = 100 MeV in the second column.  The ${\rm
p}_p/W$ is chosen as  0.1/4.814, 1/5.147 and 3/6.341~GeV. For each example
choice of ${\rm p}_p/W$, the figures represent the Dalitz plot
$d\sigma/dm_{p\bar{D}^0}dm_{pD^0}$ in the $m_{p\bar{D}^0}-m_{pD^0}$ plane in a
bin of $\mu$b/0.002$\times$0.002~GeV (upper) and invariant mass spectrum
$d\sigma/dm_{p\bar{D}^0}$ against $m_{pD^0}$ in a bin of $\mu$b/0.002 GeV (lower). The results are obtained with $10^{11}$ simulations.}\label{Dalitz2}
\end{figure}

The $X(3872)$ has a very small binding energy. Here, different binding energies
are adopted to discuss the effect of the binding energy on the Dalitz plot and
invariant mass spectrum. The plots from the first to the third column in
Fig.~\ref{Dalitz1} are for binding energies 0.1, 1, and 10 MeV, respectively (the
results of 100 MeV are shown in the second column of Fig.~\ref{Dalitz2} and
will be discussed later). An obvious concentration of events can be found for the
small binding energy 0.1 MeV in the first column. For ${\rm p}_p/W$ of 1/4.814
and 3/6.341~GeV, sharper strips can be found at $m_{p\bar{D}^{0}}$ at about 3.2 and
4.2~GeV, respectively. The invariant mass spectrum against $m_{p\bar{D}^0}$ also
exhibits a very sharp peak. Such concentration of events is from the mechanism of
collision as shown in Fig.~\ref{diagram} (a). The peak against  invariant mass
$m_{p\bar{D}^0}$ is due to the quasifree $D^0$ meson in the $X(3872)$. The
energy from the incoming proton and the transition of the $\bar{D}^{*0}$ to
$\bar{D}^0$ meson is mostly carried by the final proton and $\bar{D}^0$ meson. The
small binding energy means a large radius, that is, the large distance and weak
attraction between two constituents.  With the increase of the binding energy,
the attraction of the constituents becomes stronger, and the radius becomes
smaller. More momentum will be transferred to the $D^0$ meson, which will be
dragged by the stronger attraction of  the $\bar{D}^{*0}$ meson. It is confirmed by
comparing the results in the first and the third columns of Fig.~\ref{Dalitz1},
where, with the increase of binding energy, the strips in the Dalitz plots become
vague, and the peaks in the invariant mass spectra become wider also. In addition,
because the radius decreases with the increase of the binding energy, the cross
section is also reduced.

The results with different incoming momenta or total invariant masses are also
presented in the first to third rows in Fig.~\ref{Dalitz1}. Generally speaking,
if the incoming proton moves fast, the quasifree $D^0$ meson will be less
affected.  It can be reflected in the Dalitz plots where the strips become more
sharp with the increase of the momentum of the incoming proton. The invariant
mass spectra have more sharps peaks (for binding energy of 0.1 GeV, the peak is
more sharp because the strips are near the edge). However, the sharp peaks can be
found for all incoming momenta. Additionally, the cross section decreases rapidly with the
increase of the incoming momentum due to shorter interaction time. 

Now we turn to the compact quark state picture. The Dalitz plots and the
invariant mass spectra with different total invariant masses
are shown in the first column of Fig.~\ref{Dalitz2}.  Because the coupling
constant $g_X$ is not determined, we choose a value of 30~GeV$^{-1}$ to scale
the invariant mass spectrum.  The result shows that, with each momentum of
incoming proton or total invariant mass, the events are almost evenly
distributed in the Dalitz plot and no strip, as in the molecular state picture can be
found in the Dalitz plot. There is also no peak or structure in the invariant
mass spectrum. If the binding energy of a molecular state is very large, the
small radius and strong attraction between the constituents will make the
behavior of such state in the reaction considered similar to a compact state.
For comparison, in the second column of Fig.~\ref{Dalitz2}, the Dalitz plots for
the $X (3872)$ as a molecular state with a large binding energy $E_B$ = 100 MeV
are also presented. As we can see, these two cases give quite similar results,
especially in the row with ${\rm p}_p/W$ = 3/6.341~GeV, which is because the hadronic
molecular state tends to a compact tetraquark state with increasing binding
energy. 

The radius is an important metric for the internal structure of a hadronic state,
which will also influence the probability of collision of the incoming proton in
our work.  As we know, the radius of a $D$ or $D^{*}$ meson is smaller than 1
fm. For a molecular state of binding energy smaller than 10 MeV, the distance
between the $D$ and $\bar{D}^{*}$ meson in the $X (3872)$ is larger than 1 fm. In this
case, the incoming proton will easily collide with the $\bar{D}^{*}$ meson, and a
$\bar{D}$ meson is produced, such as shown in Fig.~\ref{Dalitz1}. If the binding
energy is larger than 10 MeV, the distance between the $D$ and $\bar{D}^{*}$ meson
in the $X (3872)$ is smaller than 1 fm, which makes the probability of collision of
the incoming proton with $\bar{D}^{*}$ in the $X (3872)$ very small. The radius also
affects the relative values of the differential cross section of different total
invariant masses $W$. Comparing the results in Fig.~\ref{Dalitz1} and the second
column of Fig.~\ref{Dalitz2}, the decrease of the invariant mass spectra
$d\sigma/dm_{pD^0}$ with the increase of $p_p/W$ becomes slower if the binding
energy becomes larger. If the $X(3872)$ is a compact state, the total cross
section even increase with the increases of the $p_p/W$ as shown in the first
column of Fig.~\ref{Dalitz2}. 

\section{Summary}\label{Summary}

In the current work, we propose a new possible reaction approach to reflect
an exotic structure of a hadronic molecular state. Taking the $X(3872)$ as an example, the
Dalitz plot and the invariant mass spectrum are estimated for $p+X(3872)\to
p+\bar{D^0}+D^0$. Two pictures of the internal structure of the $X(3872)$, the molecular
state and  compact multiquark state, are considered in the calculation. In the
molecular state picture, obvious strips and peaks can be observed in the Dalitz
plot and invariant mass spectrum, respectively.  With the increase of 
binding energy, the strips and peaks vanish gradually, and the Dalitz plot and
invariant mass spectrum tend to these in the compact multiquark picture.

The strips and  peaks are obviously from the internal structure of the exotic
state. As a molecular state, the large radius and two-constituent structure
makes the incoming proton only attack on one constituent and the other one remains
almost unaffected. However, a compact binding state of quarks, either a
quark-antiquark pair or tetraquark, the exchanged light meson should affect the
state in the whole, and no obvious strip and peak can be produced.  Though in the
current work we do not consider the explicit mechanism of the reaction of the
light meson with the compact state, the conclusion should be unaffected with
different explicit models. Such conclusion can be further confirmed with a
calculation of an unphysical large binding energy as 100~MeV, which
corresponds to a very small radius.  

Although the direct collision is unrealistic due to the lack of the exotic state
target or beam, we suggest observing such phenomenon in the production of an exotic state
in a nucleon-rich environment, such as at LHCb and PANDA. The produced exotic
state will interact with the surrounding nucleons and decay into two final
particles, for example, $D^0$ and $\bar{D}^0$ mesons here, combined with a
proton.  With different total invariant masses, the strips in Dalitz plots and the
peaks in invariant mass spectra appear in different invariant masses of two final
particles. Such proposal can provide a more determinative confirmation of the
molecular state structure of an exotic state and exclude the compact multiquark
assignment.

\section*{Acknowledgements}

This work is supported by the China National Funds for Distinguished Young
Scientists under Grant No. 11825503, the National Key Research and Development
Program of China under Contract No. 2020YFA0406400, the 111 Project under Grant
No. B20063, the National Natural Science Foundation of China under Grants No.
12247101, No. 12175091, No. 11965016, No. 11775050, No. 11775050, and No.
11675228, the CAS Interdisciplinary Innovation Team, the Fundamental Research
Funds for the Central Universities under Grant No. lzujbky-2021-sp24, and the project for top-notch innovative talents of Gansu Province.


\begin{thebibliography}{90}

%\cite{ParticleDataGroup:2018ovx}
\bibitem{ParticleDataGroup:2018ovx}
M.~Tanabashi \textit{et al.} [Particle Data Group],
``Review of Particle Physics,''
Phys. Rev. D \textbf{98}, no.3, 030001 (2018)
%doi:10.1103/PhysRevD.98.030001
%8147 citations counted in INSPIRE as of 20 Dec 2022

%\cite{Godfrey:1985xj}
\bibitem{Godfrey:1985xj}
S.~Godfrey and N.~Isgur,
``Mesons in a Relativized Quark Model with Chromodynamics,''
Phys. Rev. D \textbf{32}, 189-231 (1985)
%oi:10.1103/PhysRevD.32.189
%3056 citations counted in INSPIRE as of 20 Dec 2022

%\cite{Capstick:1986ter}
\bibitem{Capstick:1986ter}
S.~Capstick and N.~Isgur,
%``Baryons in a relativized quark model with chromodynamics,''
Phys. Rev. D \textbf{34}, no.9, 2809-2835 (1986)
%doi:10.1103/physrevd.34.2809
%1405 citations counted in INSPIRE as of 20 Dec 2022


%\cite{Chen:2016qju}
\bibitem{Chen:2016qju}
H.~X.~Chen, W.~Chen, X.~Liu and S.~L.~Zhu,
``The hidden-charm pentaquark and tetraquark states,''
Phys. Rept. \textbf{639}, 1-121 (2016)
%doi:10.1016/j.physrep.2016.05.004
%[arXiv:1601.02092 [hep-ph]].
%895 citations counted in INSPIRE as of 20 Dec 2022

%\cite{Guo:2017jvc}
\bibitem{Guo:2017jvc}
F.~K.~Guo, C.~Hanhart, U.~G.~Mei\ss{}ner, Q.~Wang, Q.~Zhao and B.~S.~Zou,
``Hadronic molecules,''
Rev. Mod. Phys. \textbf{90}, no.1, 015004 (2018)
[erratum: Rev. Mod. Phys. \textbf{94}, no.2, 029901 (2022)]
%doi:10.1103/RevModPhys.90.015004
%[arXiv:1705.00141 [hep-ph]].
%851 citations counted in INSPIRE as of 20 Dec 2022


%\cite{Liu:2019zoy}
\bibitem{Liu:2019zoy}
Y.~R.~Liu, H.~X.~Chen, W.~Chen, X.~Liu and S.~L.~Zhu,
``Pentaquark and Tetraquark states,''
Prog. Part. Nucl. Phys. \textbf{107}, 237-320 (2019)
%doi:10.1016/j.ppnp.2019.04.003
%[arXiv:1903.11976 [hep-ph]].
%419 citations counted in INSPIRE as of 20 Dec 2022


%\cite{Badalian:2012jz}
\bibitem{Badalian:2012jz}
A.~M.~Badalian, V.~D.~Orlovsky, Y.~A.~Simonov and B.~L.~G.~Bakker,
``The ratio of decay widths of X(3872) to $ \psi^{\prime}\gamma $ and $ J/\psi\gamma$ as a test of the X(3872) dynamical structure,''
Phys. Rev. D \textbf{85} (2012), 114002
%doi:10.1103/PhysRevD.85.114002
%[arXiv:1202.4882 [hep-ph]].
%33 citations counted in INSPIRE as of 24 Oct 2022

%\cite{Wang:2010ej}
\bibitem{Wang:2010ej}
T.~H.~Wang and G.~L.~Wang,
``Radiative E1 decays of X(3872),''
Phys. Lett. B \textbf{697} (2011), 233-237
%doi:10.1016/j.physletb.2011.02.014
%[arXiv:1006.3363 [hep-ph]].
%39 citations counted in INSPIRE as of 24 Oct 2022


%\cite{Swanson:2003tb}
\bibitem{Swanson:2003tb}
E.~S.~Swanson,
``Short range structure in the X(3872),''
Phys. Lett. B \textbf{588} (2004), 189-195
%doi:10.1016/j.physletb.2004.03.033
%[arXiv:hep-ph/0311229 [hep-ph]].
%545 citations counted in INSPIRE as of 24 Oct 2022

%\cite{Belle:2003nnu}
\bibitem{Belle:2003nnu}
S.~K.~Choi \textit{et al.} [Belle],
``Observation of a narrow charmonium-like state in exclusive $B^\pm \to K^\pm \pi^+ \pi^- J/\psi$ decays,''
Phys. Rev. Lett. \textbf{91} (2003), 262001
%doi:10.1103/PhysRevLett.91.262001
%[arXiv:hep-ex/0309032 [hep-ex]].
%2206 citations counted in INSPIRE as of 24 Oct 2022


%\cite{Tornqvist:2004qy}
\bibitem{Tornqvist:2004qy}
N.~A.~Tornqvist,
``Isospin breaking of the narrow charmonium state of Belle at 3872-MeV as a deuson,''
Phys. Lett. B \textbf{590} (2004), 209-215
%doi:10.1016/j.physletb.2004.03.077
%[arXiv:hep-ph/0402237 [hep-ph]].
%563 citations counted in INSPIRE as of 24 Oct 2022


%\cite{Esposito:2021vhu}
\bibitem{Esposito:2021vhu}
A.~Esposito, L.~Maiani, A.~Pilloni, A.~D.~Polosa and V.~Riquer,
``From the line shape of the X(3872) to its structure,''
Phys. Rev. D \textbf{105} (2022) no.3, L031503
%doi:10.1103/PhysRevD.105.L031503
%[arXiv:2108.11413 [hep-ph]].
%20 citations counted in INSPIRE as of 24 Oct 2022



%\cite{LHCb:2020sey}
\bibitem{LHCb:2020sey}
R.~Aaij \textit{et al.} [LHCb],
``Observation of Multiplicity Dependent Prompt $\chi_{c1}(3872)$ and $\psi(2S)$ Production in $pp$ Collisions,''
Phys. Rev. Lett. \textbf{126} (2021) no.9, 092001
%doi:10.1103/PhysRevLett.126.092001
%[arXiv:2009.06619 [hep-ex]].
%31 citations counted in INSPIRE as of 24 Oct 2022





%\cite{Barnes:2005pb}
\bibitem{Barnes:2005pb}
T.~Barnes, S.~Godfrey and E.~S.~Swanson,
``Higher charmonia,''
Phys. Rev. D \textbf{72} (2005), 054026
%doi:10.1103/PhysRevD.72.054026
%[arXiv:hep-ph/0505002 [hep-ph]].
%714 citations counted in INSPIRE as of 24 Oct 2022

%\cite{Seth:2004zb}
\bibitem{Seth:2004zb}
K.~K.~Seth,
``An Alternative Interpretation of X(3872),''
Phys. Lett. B \textbf{612} (2005), 1-4
%doi:10.1016/j.physletb.2005.02.057
%[arXiv:hep-ph/0411122 [hep-ph]].
%65 citations counted in INSPIRE as of 24 Oct 2022

%\cite{LHCb:2013kgk}
\bibitem{LHCb:2013kgk}
R.~Aaij \textit{et al.} [LHCb],
``Determination of the X(3872) meson quantum numbers,''
Phys. Rev. Lett. \textbf{110} (2013), 222001
%doi:10.1103/PhysRevLett.110.222001
%[arXiv:1302.6269 [hep-ex]].
%459 citations counted in INSPIRE as of 24 Oct 2022

%\cite{He:2022rta}
\bibitem{He:2022rta}
J.~He and X.~Liu,
``The quasi-fission phenomenon of double charm $T_{cc}^+$ induced by nucleon,''
Eur. Phys. J. C \textbf{82} (2022) no.4, 387
%doi:10.1140/epjc/s10052-022-10363-4
%[arXiv:2202.07248 [hep-ph]].
%8 citations counted in INSPIRE as of 24 Oct 2022

%\cite{He:2021smz}
\bibitem{He:2021smz}
J.~He, D.~Y.~Chen, Z.~W.~Liu and X.~Liu,
``Induced Fission-Like Process of Hadronic Molecular States,''
Chin. Phys. Lett. \textbf{39} (2022) no.9, 091401
%doi:10.1088/0256-307X/39/9/091401
%[arXiv:2109.14395 [hep-ph]].
%2 citations counted in INSPIRE as of 24 Oct 2022


%\cite{Liu:2008fh}
\bibitem{Liu:2008fh}
Y.~R.~Liu, X.~Liu, W.~Z.~Deng and S.~L.~Zhu,
``Is $X(3872) $ Really a Molecular State?,''
Eur. Phys. J. C \textbf{56} (2008), 63-73
%doi:10.1140/epjc/s10052-008-0640-4
%[arXiv:0801.3540 [hep-ph]].
%162 citations counted in INSPIRE as of 24 Oct 2022


%\cite{Machleidt:2000ge}
\bibitem{Machleidt:2000ge}
R.~Machleidt,
``The High precision, charge dependent Bonn nucleon-nucleon potential (CD-Bonn),''
Phys. Rev. C \textbf{63}, 024001 (2001)
%doi:10.1103/PhysRevC.63.024001
%[arXiv:nucl-th/0006014 [nucl-th]].
%1607 citations counted in INSPIRE as of 13 Jan 2022

%\cite{Cao:2010km}
\bibitem{Cao:2010km}
X.~Cao, B.~S.~Zou and H.~S.~Xu,
``Phenomenological analysis of the double pion production in nucleon-nucleon collisions up to 2.2 GeV,''
Phys. Rev. C \textbf{81}, 065201 (2010)
%doi:10.1103/PhysRevC.81.065201
%[arXiv:1004.0140 [nucl-th]].
%73 citations counted in INSPIRE as of 13 Jan 2022

%\cite{Tsushima:1998jz}
\bibitem{Tsushima:1998jz}
K.~Tsushima, A.~Sibirtsev, A.~W.~Thomas and G.~Q.~Li,
``Resonance model study of kaon production in baryon baryon reactions for heavy ion collisions,''
Phys. Rev. C \textbf{59}, 369-387 (1999)
[erratum: Phys. Rev. C \textbf{61}, 029903 (2000)]
%doi:10.1103/PhysRevC.59.369
%[arXiv:nucl-th/9801063 [nucl-th]].
%153 citations counted in INSPIRE as of 13 Jan 2022

%\cite{Engel:1996ic}
\bibitem{Engel:1996ic}
A.~Engel, A.~K.~Dutt-Mazumder, R.~Shyam and U.~Mosel,
``Pion production in proton proton collisions in a covariant one boson exchange model,''
Nucl. Phys. A \textbf{603}, 387-414 (1996)
%doi:10.1016/0375-9474(96)80008-F
%[arXiv:nucl-th/9601026 [nucl-th]].
%46 citations counted in INSPIRE as of 13 Jan 2022



%\cite{Casalbuoni:1996pg}
\bibitem{Casalbuoni:1996pg}
R.~Casalbuoni, A.~Deandrea, N.~Di Bartolomeo, R.~Gatto, F.~Feruglio and G.~Nardulli,
``Phenomenology of heavy meson chiral Lagrangians,''
Phys. Rept. \textbf{281} (1997), 145-238
%doi:10.1016/S0370-1573(96)00027-0
%[arXiv:hep-ph/9605342 [hep-ph]].
%626 citations counted in INSPIRE as of 24 Oct 2022

%\cite{Chen:2019asm}
\bibitem{Chen:2019asm}
R.~Chen, Z.~F.~Sun, X.~Liu and S.~L.~Zhu,
``Strong LHCb evidence supporting the existence of the hidden-charm molecular pentaquarks,''
Phys. Rev. D \textbf{100}, no.1, 011502 (2019)
%doi:10.1103/PhysRevD.100.011502
%[arXiv:1903.11013 [hep-ph]].
%79 citations counted in INSPIRE as of 29 Sep 2020


 %\cite{Voloshin:2003nt}
 \bibitem{Voloshin:2003nt}
 M.~B.~Voloshin,
 ``Interference and binding effects in decays of possible molecular component of X(3872),''
 Phys. Lett. B \textbf{579}, 316-320 (2004)
 %doi:10.1016/j.physletb.2003.11.014
 %[arXiv:hep-ph/0309307 [hep-ph]].
 %290 citations counted in INSPIRE as of 08 Aug 2021




\end{thebibliography}
\end{document}